\documentclass[sigconf,natbib=true]{acmart}
\AtBeginDocument{%
  \providecommand\BibTeX{{%
    \normalfont B\kern-0.5em{\scshape i\kern-0.25em b}\kern-0.8em\TeX}}}

\setcopyright{acmcopyright}
\copyrightyear{2024}
\acmYear{2024}
\setcopyright{acmlicensed}\acmConference[xx'18]{Proceedings of xxx}{June 03--05, 2018}{Woodstock, NY}
\acmBooktitle{Proceedings of xx (xx '18), June 03--05, 2018, Woodstock, NY}
\acmDOI{10.1145/nnnnnnn.nnnnnnn}
\acmISBN{978-1-4503-XXXX-X/18/06}

\usepackage{comment}
\usepackage{balance}
\usepackage{enumitem}
\usepackage{booktabs}
\usepackage{pifont}
\usepackage{soul}
\usepackage{amsmath}
\usepackage{tcolorbox}
\usepackage[ruled, vlined, linesnumbered]{algorithm2e}
\usepackage{subfigure}
\usepackage{multirow}

\usepackage{pgfplots}
\pgfplotsset{compat=1.16}
\usetikzlibrary{matrix}
\usetikzlibrary{patterns}
\usepgfplotslibrary{groupplots}
\pgfplotsset{compat=newest}
\usepgfplotslibrary{external}
\usepgfplotslibrary{statistics}    

\newtcolorbox{mybox}[1]{colback=orange!5!white,colframe=orange!75!black,fonttitle=\bfseries,title=#1}

\definecolor{1}{RGB}{141,211,199}
\definecolor{2}{RGB}{255,255,179}
\definecolor{3}{RGB}{190,186,218}
\definecolor{44}{RGB}{251,128,114}
\definecolor{5}{RGB}{156,209,255}
\definecolor{66}{RGB}{253,180,98}
\definecolor{77}{RGB}{179,222,105}
\definecolor{8}{RGB}{251,154,153}
\definecolor{99}{RGB}{31,120,180}
\definecolor{6}{RGB}{107,94,139}
\definecolor{9}{RGB}{78,129,121}
\definecolor{7}{RGB}{205,135,113}
\definecolor{4}{RGB}{127,66,82}

\begin{document}

\title{Adaptive In-Context Learning with Large Language Models for Bundle Generation 
}

\renewcommand{\shorttitle}{Adaptive In-Context Learning for Bundle Generation}
\author{Zhu Sun}
\affiliation{%
 \institution{A*STAR Centre for Frontier AI Research; Singapore University of Technology and Design}
 \city{Singapore}
\country{Singapore}}

\author{Kaidong Feng}
\authornote{corresponding author}
\affiliation{%
  \institution{Yanshan University}
  \city{Qinhuangdao}
  \country{China}
}
\email{kaidong3762@gmail.com}

\author{Jie Yang}
\affiliation{%
  \institution{Delft University of Technology}
  \city{Delft}
  \country{the Netherlands}
}

\author{Xinghua Qu}
\affiliation{%
\institution{Shanda AI-Lab; Tianqiao and Chrissy Chen Institute}
\city{Singapore}
\country{Singapore}
}

\author{Hui Fang}
\affiliation{%
  \institution{Shanghai University of Finance and Economics}
  \city{Shanghai}
  \country{China}
}

\author{Yew-Soon Ong}
\affiliation{%
 \institution{A*STAR Centre for Frontier AI Research; Nanyang Technological University}
\city{Singapore}
\country{Singapore}}

\author{Wenyuan Liu}
\affiliation{%
  \institution{Yanshan University}
  \city{Qinhuangdao}
  \country{China}
}
\renewcommand{\shortauthors}{Zhu Sun et al.}

\begin{abstract}
Most existing bundle generation approaches fall short in generating fixed-size bundles. Furthermore, they often neglect the underlying user intents reflected by the bundles in the generation process, resulting in less intelligible bundles. This paper addresses these limitations through the exploration of two interrelated tasks, i.e., personalized bundle generation and the underlying intent inference, based on different user sessions. Inspired by the reasoning capabilities of large language models (LLMs), we propose an adaptive in-context learning paradigm, which allows LLMs to draw tailored lessons from related sessions as demonstrations, enhancing the performance on target sessions. Specifically, we first employ retrieval augmented generation to identify nearest neighbor sessions, and then carefully design prompts to guide LLMs in executing both tasks on these neighbor sessions. To tackle reliability and hallucination challenges, we further introduce (1) a self-correction strategy promoting mutual improvements of the two tasks without supervision signals and (2) an auto-feedback mechanism for adaptive supervision based on the distinct mistakes made by LLMs on different neighbor sessions. Thereby, the target session can gain customized lessons for improved performance by observing the demonstrations of its neighbor sessions. Experiments on three real-world datasets demonstrate the effectiveness of our proposed method. 
\end{abstract}

\ccsdesc[500]{Information systems~Recommender systems}
\ccsdesc[500]{Computing methodologies~Neural networks}
\keywords{Recommendation, Bundle Generation, User Intent Inference, Large Language Models, In-Context Learning
}

\maketitle

\section{Introduction}

\begin{figure}[t]
    \centering
    \includegraphics[width=0.7\linewidth]{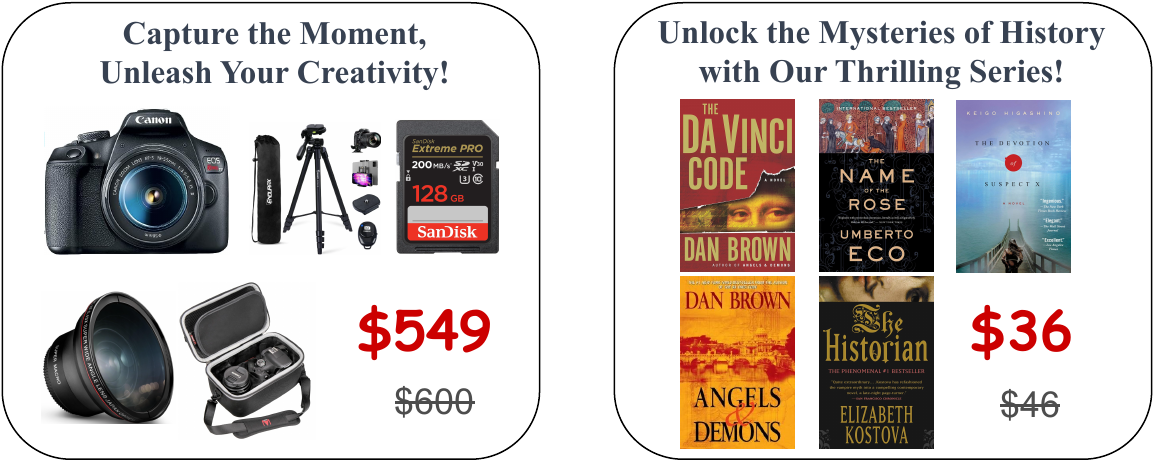}
    \vspace{-0.1in}
    \caption{Example bundles for (1) a camera and its accessories; and (2) mystery, thriller, and historical fiction.}
    \label{fig:bundle-example}
    \vspace{-0.15in}
\end{figure}

Product bundling has evolved into a crucial marketing strategy for promoting products, catering to both physical and online retailers~\cite{sar2016beyond,xie2010breaking}.  
A bundle refers to a group of products recommended or sold together as a package. These products are bundled together due to various reasons, e.g., having complementary or alternative relationships~\cite{avny2022bruce,chang2020bundle}. For instance, as depicted in Figure~\ref{fig:bundle-example}, if a customer is shopping for a camera, a bundle recommendation may include not only the camera itself but also accessories like lenses, camera bags, tripods, and memory cards - all packaged together at a discounted price. 
Therefore, product bundling can offer a beneficial solution for both customers and businesses. On the one hand, it facilitates the discovery of new items, prevents the formation of filter bubbles, and presents opportunities for potential promotions, ultimately enhancing the long-term customer experience. On the other hand, it can significantly increase product sales and drive business revenue, promoting overall economic growth within societies~\cite{sun2022revisiting}.

Given the substantial benefit, a growing body of research can be found on exploring product bundling. Most of the studies assume the pre-existence of bundles and directly dive into downstream tasks, e.g., bundle recommendation. 
In particular, 
they take co-consumed products~\cite{liu2017modeling} or user-generated lists~\cite{he2019hierarchical,chang2020bundle,he2020consistency} as synthetic bundles, or rely on manually pre-defined bundles by retailers~\cite{liu2011personalized,ge2014cost,deng2020personalized,ma2022crosscbr}. 
However, the co-consumed products
may not always reflect common intents; user-generated lists are generally limited to specific
domains (e.g. music and books); and pre-defined bundles are restricted by quantity and diversity due to the high cost of producing such bundles~\cite{sun2022revisiting}. 

Recognizing the demand for high-quality bundles, some research endeavors to develop methods for bundle generation. Early works create bundles by specifying hard constraints. The generated bundles may (1) possess a limited budget or maximum savings/customer adoption/expected revenue~\cite{xie2010breaking,xie2014generating,garfinkel2006design,yang2012bundle,beladev2016recommender,zhu2014bundle}; or (2) contain compatible products in related categories, style, or functionality~\cite{kouki2019product}.
Recently, deep learning (DL) based methods~\cite{bai2019personalized,chang2021bundle,wei2022towards} have emerged to learn the latent association of products for bundle generation. Nevertheless, these methods exhibit certain limitations, for instance, they can only form either fixed-size bundles or are limited to producing a single bundle for each user. Most importantly, they overlook the requirement for a consistent user intent underlying all products in the same bundle: products in a high-quality bundle should all reflect the same user intent that is semantically interpretable, describing a consistent user need (or purpose) in interacting with products (e.g., camera accessories, clothes for parties, or movies for Friday nights). Without such a constraint of intents, the generated bundles become less relevant and intelligible to users, thereby failing to meet their actual needs in applications.


To fill the gap, we propose simultaneously performing two interrelated tasks, i.e., \textit{generating personalized bundles} and \textit{inferring underlying intents} from user sessions\footnote{A user session is a sequence of actions  (e.g., clicks, purchases) performed by a user on a platform or website with the products during a short period (e.g., a single visit)~\cite{wang2019modeling,wu2023generic}.}. This is motivated 
by the fact that users are inclined to explore relevant products, either as alternatives or complements, based on their specific intents during a session~\cite{wang2019modeling,li2017neural}. In doing so, (1) user sessions can serve as valuable sources to create high-quality and personalized bundles; (2) the inferred user intents can enhance the interpretability of bundles, just as a well-defined bundle can clearly reflect the user's intent. However, performing both tasks at the semantic level poses significant challenges as it entails comprehending various potential motivations and contexts behind user actions and preferences, 
which can be intricate 
and ever-evolving.


To tackle the above challenges, we devise an adaptive in-context learning (AICL) paradigm leveraging the advanced reasoning capabilities of large language models (LLMs)\footnote{We use GPT-3.5-turbo in our study without further statement.}. 
This paradigm empowers LLMs to draw tailored lessons from closely related tasks, using them as demonstrations while tackling the target task.
Specifically, we first adopt the retrieval augmented generation~\cite{jiang2023active} to identify the nearest neighbor sessions for each target session, and then create prompts to instruct LLMs to perform both tasks in neighbor sessions. To enhance reliability and mitigate the hallucination issue, we further develop (1) a self-correction strategy to foster mutual improvements in both tasks without 
supervision signals; and (2) an auto-feedback mechanism to recurrently offer adaptive supervision by comparing LLMs' output and the labels. Subsequently, we guide LLMs to provide a summary of rules derived from the entire task execution process to prevent recurring errors in the future. 
Finally, the two tasks in the target session are performed by observing demonstrations of its neighbor sessions.
Different neighbors may possess distinct mistakes made by LLMs, thereby receiving different feedback. It thus enables LLMs to seek adaptive and customized lessons for improved performance on the target session.

Our contributions are three-fold. 
\textbf{First}, we propose a new research question to perform two interrelated tasks, i.e., bundle generation and intent inference, based on user sessions. As such, the generated bundles are more intelligible and aligned with users' actual needs. \textbf{Second}, we design a novel adaptive in-context learning paradigm for our defined tasks, which enables LLMs to seek tailored lessons from neighbor sessions as demonstrations. To achieve this, we devise step-by-step strategies evolving from mutual self-correction (self-supervision) to adaptive auto-feedback (auto external-supervision), and finally rules summarization (self-supervision). This is a novel idea in the context of using LLMs for recommendation.
\textbf{Lastly}, we conduct experiments on three public datasets. The results show that AICL surpasses baselines on the bundle generation task, and the inferred intents are of high quality, comparable to or even exceeding those annotated by humans.

\section{Related Work}

\subsection{Recommendation with Prebuilt Bundles}
Many works assume the presence of prebuilt bundles and immediately delve into the downstream task of bundle recommendation. Early methods generate bundles by satisfying certain constraints, e.g., limited cost~\cite{liu2011personalized,ge2014cost}. 
Later, factorization-based methods decompose user-item and user-bundle interactions to learn users' interests over items and bundles, respectively~\cite{liu2014recommending,cao2017embedding}. Recently, DL-based methods (e.g., DAM~\cite{chen2019matching}, AttList~\cite{he2019hierarchical}, CAR~\cite{he2020consistency}, BRUCE~\cite{avny2022bruce}, and BundleGT~\cite{wei2023strategy}) adopt the attention mechanism to learn item-bundle affinity and user-bundle preference. Other methods adopt graph or hypergraph convolutional networks to better infer users' preference towards bundles, such as BGCN~\cite{chang2020bundle}, BundleNet~\cite{deng2020personalized},
CrossCBR~\cite{ma2022crosscbr},
MIDGN~\cite{zhao2022multi},
UHBR~\cite{yu2022unifying}, SUGER~\cite{zhang2022suger}, and DGMAE~\cite{ren2023distillation}.
However, these methods are all based on prebuilt bundles, i.e., either co-consumed products, user-generated lists, or predefined ones by retailers as summarized in Table~\ref{tab:prebuilt-bundles}. 
They ignore the fact that (1) co-consumed products may not consistently represent shared intentions; (2) user-generated lists are typically confined to specific domains (e.g., music and books); 
and (3) pre-defined bundles are constrained by their limited quantity and diversity, primarily due to the high production costs associated with them.  

\begin{table*}[!t]
    \footnotesize
    \addtolength{\tabcolsep}{0.9pt}
    \caption{Approaches with bundle generation. `?' means the answer is not found based on the paper and source code if available.}
    \label{tab:bundle-generation}
    \vspace{-0.15in}
    \begin{tabular}{c|ccccccccccccccccccc}
    \toprule
    &\cite{agrawal1994fast}
    &\cite{sun2022revisiting}
    &\cite{garfinkel2006design}
    &\cite{xie2010breaking}
    &\cite{pathak2017generating}
    &\cite{parameswaran2011recommendation}
    &\cite{yang2012bundle}
    &\cite{beladev2016recommender}
    &\cite{xie2014generating}
    &\cite{dragone2018no}
    &\cite{zhu2014bundle}
    &\cite{kouki2019product}
    &\cite{bai2019personalized}
    &\cite{chen2019pog}
    &\cite{wei2022towards}
    &\cite{deng2021build}
    &\cite{he2022bundle}
    &\cite{chang2021bundle}
    &Ours\\\midrule
    {Dynamic Bundle Size}
    &\ding{56}&\ding{56}&\ding{52}&\ding{52}&\ding{52}&\ding{52}&\ding{56}&\ding{56}&\ding{56}&\ding{52}&\ding{56}&\ding{56}&\ding{56}&\ding{52}&?&\ding{56}&\ding{52}&\ding{52}&\ding{52}\\
    {Multiple Bundles} &\ding{52}&\ding{52}&\ding{56}&\ding{52}&\ding{56}&\ding{56}&\ding{52}&\ding{52}&\ding{52}&\ding{56}&\ding{56}&\ding{56}&\ding{52}&\ding{56}&\ding{56}&\ding{56}&\ding{56}&\ding{52}&\ding{52}\\
    {Personalized Bundles} &\ding{56}&\ding{56}&\ding{52}&\ding{52}&\ding{52}&\ding{52}&\ding{52}&\ding{52}&\ding{52}&\ding{52}&\ding{52}&\ding{56}&\ding{52}&\ding{52}&\ding{52}&\ding{52}&\ding{52}&\ding{52}&\ding{52}\\
    {Intent Inference} &\ding{56}&\ding{56}&\ding{56}&\ding{56}&\ding{56}&\ding{56}&\ding{56}&\ding{56}&\ding{56}&\ding{56}&\ding{56}&\ding{56}&\ding{56}&\ding{56}&\ding{52}&\ding{56}&\ding{56}&\ding{56}&\ding{52}\\
    \bottomrule
    \end{tabular}
\end{table*}

\begin{table}[t]
    \centering
    \footnotesize
    \addtolength{\tabcolsep}{5pt}
    \vspace{-0.1in}
    \caption{Approaches with different prebuilt bundles.}\label{tab:prebuilt-bundles}
    \vspace{-0.15in}
    \begin{tabular}{c|c}
    \toprule
    Type & Methods \\\midrule
    Co-consumed Products &  \cite{liu2017modeling} \\
    User Generated Lists & \cite{liu2014recommending,cao2017embedding,chen2019matching,he2019hierarchical,chang2020bundle, he2020consistency,deng2020personalized,zhao2022multi,ma2022crosscbr,avny2022bruce,yu2022unifying,zhang2022suger,wei2023strategy,ren2023distillation} \\
    Predefined by Retailers & \cite{liu2011personalized,ge2014cost,deng2020personalized,ma2022crosscbr,avny2022bruce,wei2023strategy,ren2023distillation} \\
    \bottomrule
    \end{tabular}
    \vspace{-0.15in}
\end{table}

\subsection{Recommendation with Bundle Generation}
Several studies explore bundle recommendation with generation. 
In the early stage, bundles are created via frequent itemsets mining algorithm~\cite{agrawal1994fast,sun2022revisiting}. 
Later, they are formed by adhering to specific hard constraints. For instance, greedy-based methods create bundles by minimizing the cost~\cite{garfinkel2006design,xie2010breaking}, or fulfilling other requirements~\cite {parameswaran2011recommendation,pathak2017generating}. Heuristic methods form bundles by maximizing customer adoption~\cite{yang2012bundle}, expected revenue~\cite{beladev2016recommender}, or sharing the same category~\cite{kouki2019product}. 
Preference elicitation methods produce bundles via users' preference for cost and quality~\cite{xie2014generating,dragone2018no}.
Others frame bundle generation as a Quadratic Knapsack Problem~\cite{zhu2014bundle} to maximize the expected reward.
Recent studies mainly resort to DL techniques for bundle generation, such as Seq2Seq based methods~\cite{bai2019personalized,chen2019pog,wei2022towards}, and graph-generation based method~\cite{chang2021bundle}.
Other works treat it as combinatorial optimization~\cite{deng2021build} or Markov Decision Process~\cite{he2022bundle} and adopt reinforcement learning to compose bundles. 
However, they suffer from various drawbacks (see Table~\ref{tab:bundle-generation}): some methods can only generate fixed-size bundles or a single bundle for each user~\cite{pathak2017generating,zhu2014bundle}; some overlook personalization~\cite{kouki2019product,agrawal1994fast}; and others exhibit high complexity and limited scalability~\cite{chang2021bundle}. Most importantly, they generally fail to understand user intent at the semantic level during bundle generation. Consequently, the created bundles may be less comprehensible and aligned with users' actual needs.

\subsection{Intent-Aware Session Recommendation}
Our proposal of simultaneously generating personalized bundles and inferring underlying intents from user sessions is related to intent-aware session recommendation~\cite{wang2019modeling,li2017neural}. Specifically, early work learns the main intent in a session to help infer user preference~\cite{li2017neural,liu2018stamp,yu2020tagnn}. However, only learning the main intent may limit the model performance, as items in a session may often reveal multiple intents. Hence, later works capture multiple intents in a session~\cite{tanjim2020attentive,wang2019modeling,li2022enhancing}. However, they can only learn a fixed number (one or multiple) of latent intents in the session, which is overly rigid and cannot faithfully unveil user intents in a session. In contrast, our study aims to generate an adaptive number of bundles and underlying intents at the semantic level based on user sessions. 
Closet to our work is the method proposed in~\cite{zhu2020sequential} that can learn multiple user intents in a session; its assumption, i.e., items belonging to the same category indicate the same intent, however, may not always hold in reality. Our method instead, allows to generate multiple intents of any type, not restricted to item categories. 

\subsection{LLMs for Recommendation}
The remarkable achievements of LLMs have led to their widespread adoption for more effective recommendation~\cite{wu2023survey,zhang2023chatgpt,harte2023leveraging}. Many works adopt \textit{in-context learning} (ICL) to align LLMs for recommendation.  For instance, 
Zhai et al.~\cite{zhai2023knowledge}  transform knowledge graphs into knowledge prompts for more
explainable recommendation. 
Other studies~\cite{dai2023uncovering,sanner2023large,he2023large} highlight ChatGPT's potential to mitigate the cold start issue and provide explainable recommendations. 
Another line of research exploits \textit{parameter-efficient fune-tuning} (PEFT) to align LLMs for recommendation, such as TallRec~\cite{bao2023tallrec}, PALR~\cite{yin2023heterogeneous}, InstructRec~\cite{zhang2023recommendation}, and 
HKFR~\cite{chen2023palr}. 
Despite the effectiveness of these LLM-based methods, they are all designed for individual item recommendation. On the contrary, our study attempts to leverage the capability of LLMs through ICL for personalized bundle 
(a set of associated items) 
generation and underlying intent reasoning. Instead of using randomly sampled few-shot examples~\cite{sanner2023large}, we employ the retrieval augmented generation to retrieve the dataset and identify the most correlated neighbor sessions. On this basis, we create demonstrations via the proposed self-correction and auto-feedback strategies. This process empowers LLMs to take customized and adaptive lessons from neighboring sessions, ultimately leading to enhanced performance in the test session. 

\subsection{Prompting Methods for LLMs}
The fact that LLMs have seen extensive application across diverse tasks and domains has also made a strong impact in the research communities, where an increasing amount of research work is being found on designing prompts for LLMs utilization.
Many advanced prompting methods have been introduced to guide LLMs in generating more specific, accurate and high-quality responses, such as Chain-of-Thought~\cite{wei2022chain}, Tree-of-Thought~\cite{yao2023tree}, Self-Consistency~\cite{wang2023self}, Self-Reflection~\cite{madaan2023self}, Generated Knowledge~\cite{liu2022generated}, Least-to-Most~\cite{zhou2023least} and Retrieval Augmentation~\cite{jiang2023active}.
These methods offer promising opportunities in recommendation contexts, yet come also with the crucial challenge of creating appropriate prompts that are tailored to specific recommendation tasks. In response to this challenge, our study introduces novel strategies specifically designed to leverage LLMs effectively for bundle generation and intent inference.

\section{The Proposed Methodology}
\begin{figure*}[t]
\centering
\includegraphics[width=1.0\linewidth]{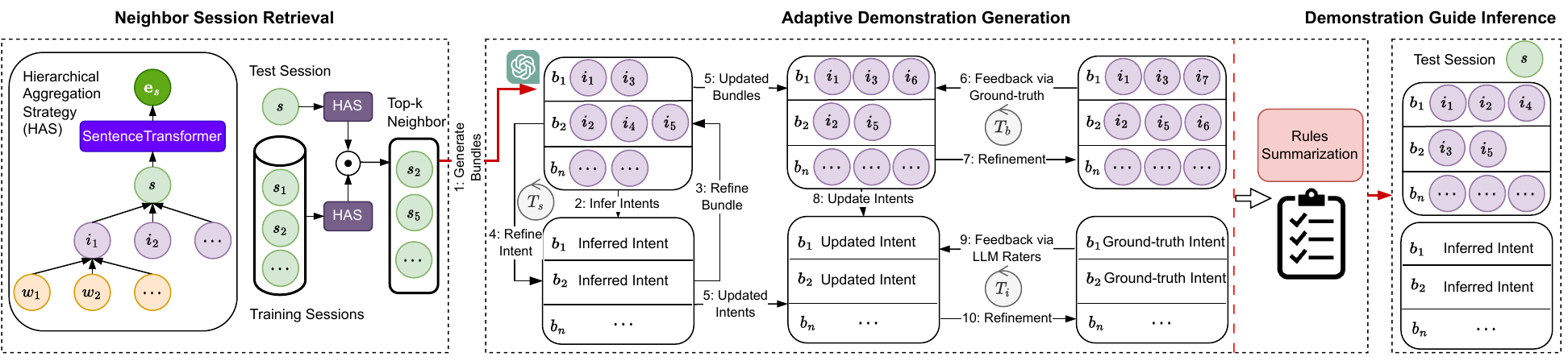}
    \vspace{-0.25in}
    \caption{The overall framework of AICL. We take one test session and its top-1 neighbor session as an example for illustration.}
    \label{fig:framework}
    \vspace{-0.1in}
\end{figure*}
We design an adaptive in-context learning (AICL) paradigm for LLMs to simultaneously perform two interrelated tasks: \textit{generating bundles} and \textit{inferring underlying intents}, from user sessions. This is motivated by the fact that users tend to explore highly correlated products, either alternatives or complements, based on their specific intents during a session~\cite{wang2019modeling}. The two tasks defined can be mutually reinforced and enhanced. Specifically, effective user intents can help identify relevant products to form improved bundles and enhance interpretability.  Meanwhile, a well-defined bundle, in turn, can provide a clearer elucidation of the user's intent.

\smallskip\noindent\textbf{Model Overview}.
Our core idea is to enable LLMs to seek tailored and adaptive lessons from closely related tasks as demonstrations while performing the target task. This is different from existing ICL-based recommendation methods~\cite{dai2023uncovering,sanner2023large}, which rely on randomly sampled examples and static instructions. Our AICL mainly consists of three modules as shown in Figure~\ref{fig:framework}. 
\begin{itemize}[leftmargin=*]
    \item \textit{Neighbor Session Retrieval} exploits the retrieval augmented generation~\cite{jiang2023active} to identify from the entire dataset the most correlated sessions for each target session regarding products contained. Such neighbor sessions will be used to generate demonstrations. 
    \item \textit{Adaptive Demonstration Generation} creates prompts to instruct LLMs to perform both tasks in neighbor sessions. To enhance reliability and mitigate the hallucination issue, we develop a self-correction strategy to foster mutual improvements in both tasks without the need for any supervision signals. Afterwards, an auto-feedback mechanism is devised to recurrently provide adaptive supervision by comparing the outputs of LLMs and the ground truth (labels). Finally, it directs LLMs to provide a summary of rules derived from the entire task execution process to prevent recurring errors in the future.
    \item \textit{Demonstration Guided Inference} observes demonstrations of neighbor sessions to perform the two tasks in the corresponding target session. Different neighbor sessions may encounter distinct mistakes/errors made by LLMs, thereby receiving diverse feedback. Hence, it empowers LLMs to seek tailored and adaptive lessons for improved performance on the target session. 
\end{itemize}
Equipped with the three modules, our AICL paradigm is capable of effectively creating multiple, personalized, and intelligible bundles with adaptive sizes given any user session. 
\subsection{Neighbor Session Retrieval (NSR)}\label{subsec:nsr}
Existing ICL-based recommendation methods~\cite{dai2023uncovering,sanner2023large,liu2023chatgpt} randomly sample few-shot examples to instruct LLMs (e.g., ChatGPT) for the inference process. However, when the sampled examples are less relevant to the target task, LLMs receive only a restricted amount of useful knowledge to guide them~\cite{huang2023enhancing}. To resolve this issue, the Neighbor Session Retrieval module (NSR) employs the retrieval augmented generation~\cite{jiang2023active} to retrieve the entire dataset and identify the most correlated sessions (i.e., nearest neighbor sessions) for each target session regarding products contained. 
By doing so, we can acquire a wealth of valuable knowledge and lessons that can significantly enhance performance.

To this end, we devise a hierarchical aggregation strategy to get the representation of each session. First,  for each item, we process its title via the natural language processing toolkit - NLTK (\url{nltk.org})
and regular expressions, to remove stop words and special characters (e.g., \&, \#), etc. The processed item title is treated as its description denoted as $i\leftarrow[w_1, w_2, \cdots]$, where $w_x$ means an individual word. 
Then, we concatenate descriptions of items within one session to represent its session description denoted as $s \leftarrow [{i_1}, {i_2}, \cdots]$.
Subsequently, we feed the session description into SentenceTransfomers (all-MiniLM-L6-v2)~\cite{reimers2019sentence} to get its latent representation.
given by, $\boldsymbol{e}_{s} = \text{SentenceTransformer} (s).$
%
%
Finally, for each target (test) session, we calculate its cosine similarity with all training sessions using the learned latent representations, to identify its top-$k$ nearest neighbors. Such neighbor sessions are used for generating demonstrations to enhance LLMs' performance on the test session.

It is noteworthy that one may consider using the embeddings (i.e., representations) generated by LLMs (e.g., OpenAI text-embedding-ada-002) to calculate the similarity. Instead, we choose SentenceTransformers due to three aspects. 
(1) \textbf{Lower Time Complexity}. The dimension of the embedding output by SentenceTransformer (all-MiniLM-L6-v2) is 384, whereas the dimension output by OpenAI (text-embedding-ada-002) is 1536. The smaller embedding size will greatly reduce the time complexity for finding the nearest neighbors.
(2) \textbf{Less API Usage Cost}. SentenceTransformer is an open-source Python framework, whereas we have to pay when using LLMs such as OpenAI (text-embedding-ada-002). In comparison, SentenceTransformer helps save much financial cost, especially for large-scale session datasets.
(3) \textbf{Comparable Performance}. SentenceTransformers can obtain comparable performance as LLMs such as OpenAI (text-embedding-ada-002), which can be verified by the experimental results in Section~\ref{subsec:ablation}.


\subsection{Adaptive Demonstration Generation (ADG)}
Next, ADG designs proper prompts to instruct LLMs to perform both tasks (i.e., bundle generation and intent reasoning) on these neighbor sessions, with the goal of creating demonstrations for improved performance on the target session. 

First, a prompt that asks LLMs to generate bundles is created as - \textcolor{black!80}{\textit{A bundle can be a set of alternative or complementary products that are purchased with a certain intent.  Please detect bundles from a sequence of products. Each bundle must contain multiple products. Here are the products and descriptions: \{[product X: title, $\dots$]\}. The answer format is: \{`bundle number':[`product number']\}. No explanation for the results}}.
As a result, LLMs will generate bundles and output them in the requested format. Subsequently, another prompt is passed to LLMs to infer the intent behind each generated bundle as - \textcolor{black!80}{\textit{Please use 3 to 5 words to generate intents behind the detected bundles, the output format is: \{`bundle number':`intent'\}}}. Note that we adopt the average number of words in the ground truth intents (i.e., 3.4) as a constraint to prevent overly long intents. To further enhance the reliability and mitigate the hallucination issue, we design the following strategies for more robust performance.  

\subsubsection{Mutual Self-Correction}
Given the generated bundles and intents, we design a self-correction strategy to foster mutual improvements in both tasks without the need for any supervision signals.  As emphasized, the two tasks are interrelated, that is, the user intent can increase the interpretability of bundles and help identify relevant products to form bundles, while the well-defined bundles can clearly reflect the user's intent. We thus create a prompt to exploit the inferred intent to refine the generated bundles - \textcolor{black!80}{\textit{Given the generated intents, adjust the detected bundles with the product descriptions. The output format is: \{`bundle number':[`product number']\}}}. 
If there is any adjustment, the adjusted bundles, in turn, are further employed to refine the intents - \textcolor{black!80}{
\textit{Given the adjusted bundles, regenerate the intents behind each bundle, the output format is: \{`bundle number':`intent'\}}
}. We repeat the above process $T_s$ times or until there is no further adjustment.

\subsubsection{Adaptive Auto-Feedback}\label{subsubsec:auto-feedback}
Despite the effectiveness of the self-correction strategy, explicit supervision is more helpful to better guide LLMs. Hence, we proceed to design an auto-feedback mechanism to recurrently provide adaptive instructions for LLMs regarding the generated bundles and intents, thereby chasing further performance enhancement. 
We start from the generated bundles. Based on the ground truth bundles and potential mistakes made by LLMs, we define five types of supervision signals: 
\begin{itemize}[leftmargin=*]
    \item Type 1: correct and should be kept;
    \item Type 2: invalid and should be removed (not containing any products in the ground truth bundles); 
    \item Type 3: containing unrelated products to be removed;
    \item Type 4: missing some products and should append other related products (bundle size>1);
    \item Type 5: missing some products and should contain at least two related products (bundle size=1).
\end{itemize}

Since multiple bundles may be generated by LLMs given a session, we calculate the Jaccard similarity between the generated bundles and ground truth bundles. Thus, we can match and compare them, and then automatically provide the corresponding supervision signals. Accordingly, we pass the prompt to LLMs to refine its generated bundles - \textcolor{black!80}{
\textit{Here are some tips for the detected bundles in your answer: \{[bundle X is Type X, $\dots$]\}; Adjust the bundles based on the tips in your answer. Please output the adjusted bundles with the format: \{`bundle number':[`product number']\}}}. 
We repeat such a process $T_b$ times or until only the Type 0 signal is returned. Since LLMs could make different mistakes for various generated bundles in different sessions, the auto-feedback mechanism can recurrently offer adaptive supervision based on the ground truth bundles. 

We now proceed with the inferred intents. We first ask LLMs to re-infer intents for the above updated bundles - \textcolor{black!80}{
\textit{Please use 3 to 5 words to generate intents behind the detected bundles, the output format is: \{`bundle number':`intent'\}}}. Afterwards, similar to the bundle auto-feedback generation, we define 3 types of supervision signals for the inferred intents as below:
\begin{itemize}[leftmargin=*]
    \item Type 1: (Naturalness) 
    be more natural;
    \item Type 2: (Coverage) 
    cover more products within the bundle;
    \item Type 3: (Motivation) 
    have a more motivational description.
\end{itemize}
We seek to compare the intents inferred by LLMs and the ground truth in three aspects. In particular, \textit{Naturalness} indicates whether the intent is easy to read and understand; \textit{Coverage} implies to what extent the items in the bundle can be covered by the intent; and \textit{Motivation} suggests whether the intent contains motivational description, i.e., describing the purpose of the bundle by activities, events, or actions. For example, the intent `assembling computer' is motivational, whilst `different computer accessories' is not. 
Based on this, we can provide the corresponding supervision signal defined above. However, it may necessitate human evaluation, potentially leading to labor-intensive tasks. To remedy this, we adopt two LLMs (another ChatGPT and Claude-2) as raters to automatically conduct the intent assessment task via the Intent Assessment Prompt demonstrated on the right side.

For the sake of robustness, we instruct each rater to repeat the evaluation process three times and calculate the average as the final rating. For each metric, we compare the rating between the generated intent and ground truth. If any rater provides a lower rating to the generated intents on any metric, we then provide the corresponding supervision signals to LLMs for further refinement via the prompt - \textcolor{black!80}{
\textit{Here are some tips for the generated intents in your answer: regenerate intent X to \{[Type X, $\cdots$]\}. Please output the regenerated intents with the format: \{`bundle number':`intent'\}}}.
We repeat the above process either $T_i$ times or until the ratings of generated intents are no lower than those of ground truth across the three metrics for both raters. 


\begin{mybox}{Prompt: \emph{Intent Assessment}}
\textit{
The intent should describe the customer’s motivation well in the purchase of the product bundles. You are asked to evaluate two intents for a bundle, using three metrics: Naturalness, Coverage, and Motivation. The details and scales of each metric are listed below. \\
Naturalness: \\
1 - the intent is difficult to read and understand \\
2 - the intent is fair to read and understand \\
3 - the intent is easy to read and understand \\
Coverage: \\
1 - only a few items in the bundle are covered by the intent \\
2 - around half items in the bundle are covered by the intent \\
3 - most items in the bundle are covered by the intent \\
Motivation: \\
1 - the intent contains no motivational description \\
2 - the intent contains motivational description \\
Following are the bundles that we ask you to evaluate: \\
\{[product X: title, $\dots$]\}, 
\{intent X, intent GT\} \\
Please answer in the following format: \{`intent number':
[`Naturalness':score, `Coverage':score, `Motivation':score]\}.}
\end{mybox}

\subsubsection{Rules Summarization}
Beyond the conversation above, we further instruct LLMs to derive useful rules from the entire task execution process to prevent recurring errors in the future, with the prompt - \textcolor{black!80}{
\textit{Based on the conversations above, which rules do you find when detecting bundles?}}. Here are examples of some generated rules: 
(1) products with similar intents are grouped together in a bundle;
(2) missing products can be appended to the bundles if they are related to the intent; 
(3) the adjusted bundles should reflect the intent and include relevant products from the sequence;
(4) the intent behind a bundle can be inferred from the combination of products and their intended use;
and (5) the regenerated intents should be descriptive and motivational.

\subsection{Demonstration Guided Inference (DGI)}
Given the constructed demonstration, we ask LLMs to perform the two tasks on the corresponding test session via the prompt - \textcolor{black!80}{
\textit{Based on the rules above, detect bundles for the below product sequence: \{[product X: title, $\dots$]\}.
The answer format is: \{`bundle number':[`product number']\}. No explanation for the results}} 
and - \textcolor{black!80}{
\textit{Please use 3 to 5
words to generate intents behind the detected bundles, the output format is \{`bundle number':`intent'\}.}} 
By observing demonstrations of neighbor sessions, LLMs can seek tailored and adaptive lessons for improved performance on the test session. 

\subsection{Complexity Discussion}
The time complexity of our method mainly comes from three modules: (1) Neighbor Session Retrieval, (2) Adaptive Demonstration Generation, and (3) Demonstration Guided Inference. 
For (1), using SentenceTransformer to obtain the session embedding is quite fast. The main complexity lies in the pairwise similarity calculation, i.e., $\mathcal{O}(\vert\mathcal{S}_r\vert\times\vert\mathcal{S}_t\vert)$, where $\vert\mathcal{S}_r\vert$ and $\vert\mathcal{S}_t\vert$ are the total number of training and test sessions, respectively. For (2), it involves calling ChatGPT API for iterative adjustment via self-correction and adaptive auto-feedback, which constitutes the bulk of the complexity. This process mainly depends on the network latency and server load. For (3), it involves doing inference with the demonstration as context using ChatGPT API, which depends on the token size of the context and factors mentioned in (2).

In real-world applications, the computation within each of the three modules can be done in parallel to speed up the whole process. Besides, during inference, for each test session, step (1) can be sped up by employing Product Quantization (PQ)~\cite{jegou2010product} to compress text embedding as quantization-based representation~\cite{zhang2021joint}, thereby reducing the cost of similarity computation. Step (2) involves multiple iterations and refinements using LLMs for generating demonstrations, which constitute the bulk of the complexity and can be done offline and stored in the database in advance. This is because the demonstrations are generated using training sessions only. We can offline generate them on all training sessions or a reasonable number of representative training sessions. In summary, our method is reasonably feasible for practical application. 
\section{Experiments and Results}
We conduct extensive experiments on three public datasets to demonstrate the effectiveness of our proposed AICL paradigm. For reproducibility~\cite{sun2022daisyrec}, our code is available at \url{https://github.com/BundleRec/bundle_generation}. 

\begin{table}[t]
    \centering
    \footnotesize
    \caption{The statistics of the three bundle datasets.}
    \label{tab:datasets}
    \addtolength{\tabcolsep}{5pt}
    \vspace{-0.15in}
    \begin{tabular}{c|c|c|c}
    \toprule
    &Electronic &Clothing &Food \\\midrule
    \#Users  &888 &965 &879\\
    \#Items &3499 &4487 &3767\\
    \#Sessions &1145 &1181 &1161\\
    \#Bundles &1750 &1910 &1784\\
    \#Intents &1422 &1466 &1156\\
    \#User-Item Interactions &6165 &6326 &6395\\ 
    \#User-Bundle Interactions &1753 &1912 &1785\\
    Average Bundle Size &3.52 &3.31 &3.58 \\
    \bottomrule
    \end{tabular}
    \vspace{-0.1in}
\end{table}
\begin{table*}[t]
\centering
\footnotesize
\addtolength{\tabcolsep}{7pt}
\caption{The performance on bundle generation. For the sake of robustness, we run each method five times to report the average results; the best results are highlighted in bold; the runner-up is underlined; and `$\dagger$' refers to our method significantly outperforms the best-performed baselines with a paired t-test ($p$-value < 0.05). 
}\label{tab:result-generation}
\vspace{-0.15in}
\begin{tabular}{c|c|c|c|c|c|c|c|c|c|c}
\toprule
\multicolumn{2}{c|}{}
&\multicolumn{3}{c|}{Electronic} &\multicolumn{3}{c|}{Clothing} &\multicolumn{3}{c}{Food}  \\\cline{3-11}
\multicolumn{2}{c|}{} &Precision&Recall&Coverage&Precision&Recall&Coverage&Precision&Recall&Coverage\\\midrule
\multirow{4}{*}{\rotatebox[origin=c]{90}{Non-LLMs}} &Freq &0.423 &0.597 &0.701 &0.532 &0.566 &0.698 &0.491 &0.525 &0.684 \\ 
&BBPR &0.260&0.122&0.433&0.239&0.211&0.449&0.210&0.183&0.416 \\ 
&POG  &0.339&0.250&0.412&0.312&0.221&0.399&0.365&0.266&0.393 \\ 
&BYOB &0.340&0.294&0.361&0.311&0.273&0.457&0.304&0.253&0.427 \\\hline 
\multirow{4}{*}{\rotatebox[origin=c]{90}{LLMs}} &T5   &0.553&0.553&0.502&0.572&0.581&0.507&0.575&0.574&0.451 \\
&Zero-shot 
&0.580 &0.820 &0.720 &\underline{0.603} &0.752 &\underline{0.788} &0.604 &0.815 &0.748 \\ 
&Few-shot  
&\underline{0.587} &\underline{0.825} &\underline{0.724} &0.595 &\textbf{0.836} &0.781 &\underline{0.647} &\underline{0.833} &\underline{0.749} \\ 
&AICL 
&\textbf{0.679}$^\dagger$ &\textbf{0.859}$^\dagger$ &\textbf{0.741}$^\dagger$ &\textbf{0.677}$^\dagger$ &\underline{0.788} &\textbf{0.839}$^\dagger$ &\textbf{0.698}$^\dagger$ &\textbf{0.851}$^\dagger$ &\textbf{0.755}$^\dagger$ \\\bottomrule
\end{tabular}
\vspace{-0.1in}  
\end{table*}
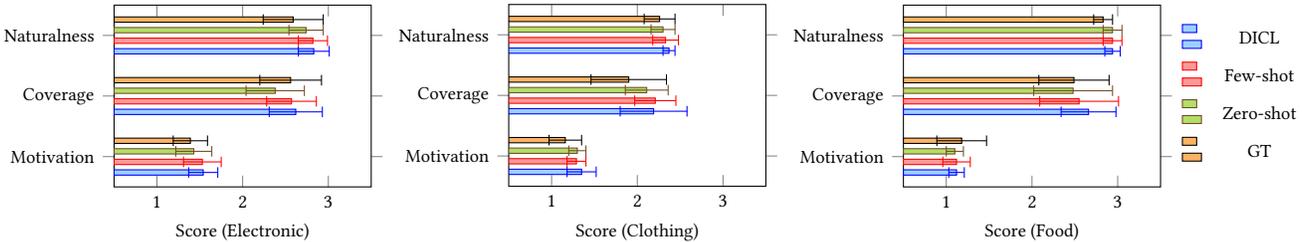
\begin{figure*}[t]
    \centering
    \subfigure
    {
    \begin{tikzpicture}
    \begin{axis}[
        xbar,
        width=4.9cm,
	height=3.7cm,  
        bar width=2pt,
        xlabel={Score (Electronic)},
        ylabel style ={font = \footnotesize},
        xlabel style ={font = \footnotesize},
        scaled ticks=false,
        tick label style={/pgf/number format/fixed, font=\footnotesize},
        xmin=0, xmax=3.5, 
        ymin=-0.5, ymax=2.5,
        xtick={0,1,2,3},
        yticklabels={Motivation, Coverage, Naturalness},
        ytick={0,1,2},
        legend style={at={(1.05,0.5)}, anchor=west, 
        minimum height=1cm,
        legend columns=1, column sep=0.2cm, draw=none, font=\footnotesize},
    ]
    \addplot+[mark=none,fill=5][error bars/.cd,x dir=both, x explicit]
        coordinates {
        (2.83,2) +- (0.18,0)
        (2.62,1) +- (0.31,0)
        (1.54,0) +- (0.17,0)
        };
    \addplot+[mark=none,fill=8][error bars/.cd,x dir=both, x explicit]
        coordinates { 
        (2.82,2) +- (0.17,0)
        (2.57,1) +- (0.29,0)
        (1.53,0) +- (0.22,0)
        };
    \addplot+[mark=none,fill=77][error bars/.cd,x dir=both, x explicit]
        coordinates { 
        (2.74,2) +- (0.2,0)
        (2.38,1) +- (0.34,0)
        (1.43,0) +- (0.21,0)
        };
    \addplot+[mark=none,fill=66][error bars/.cd,x dir=both, x explicit]
        coordinates {
        (2.59,2) +- (0.35,0)
        (2.56,1) +- (0.36,0)
        (1.39,0) +- (0.2,0)
        };    
    \end{axis}
    \end{tikzpicture}} 
    \hspace{-0.05in}
    \subfigure
    {
    \begin{tikzpicture}
    \begin{axis}[
        xbar,
        width=4.9cm,
		height=3.7cm,
        bar width=2pt,
        xlabel={Score (Clothing)},
        ylabel style ={font = \footnotesize},
        xlabel style ={font = \footnotesize},
        scaled ticks=false,
        tick label style={/pgf/number format/fixed, font=\footnotesize},
        xmin=0, xmax=3.5, 
        ymin=-0.5, ymax=2.5,
        xtick={0,1,2,3},
        yticklabels={Motivation, Coverage, Naturalness},
        ytick={0,1,2},
        legend style={at={(1.05,0.5)}, anchor=west, 
        minimum height=1cm,
        legend columns=1, column sep=0.2cm, draw=none, font=\footnotesize},
    ]
    \addplot+[mark=none,fill=5][error bars/.cd,x dir=both, x explicit]
        coordinates {
        (2.37,2) +- (0.07,0)
        (2.19,1) +- (0.39,0)
        (1.35,0) +- (0.17,0)
        };
    \addplot+[mark=none,fill=8][error bars/.cd,x dir=both, x explicit]
        coordinates { 
        (2.33,2) +- (0.15,0)
        (2.21,1) +- (0.24,0)
        (1.29,0) +- (0.11,0)
        };
    \addplot+[mark=none,fill=77][error bars/.cd,x dir=both, x explicit]
        coordinates { 
        (2.30,2) +- (0.14,0)
        (2.11,1) +- (0.25,0)
        (1.30,0) +- (0.10,0)
        };
    \addplot+[mark=none,fill=66][error bars/.cd,x dir=both, x explicit]
        coordinates {
        (2.26,2) +- (0.18,0)
        (1.90,1) +- (0.44,0)
        (1.16,0) +- (0.19,0)
        };    
    \end{axis}
    \end{tikzpicture}} 
    \hspace{-0.05in}
    \subfigure
    {
    \begin{tikzpicture}
    \begin{axis}[
        xbar,
        width=4.9cm,
		height=3.7cm,
        bar width=2pt,
        xlabel={Score (Food)},
        ylabel style ={font = \footnotesize},
        xlabel style ={font = \footnotesize},
        scaled ticks=false,
        tick label style={/pgf/number format/fixed, font=\footnotesize},
        xmin=0, xmax=3.5, 
        ymin=-0.5, ymax=2.5,
        xtick={0,1,2,3},
        yticklabels={Motivation, Coverage, Naturalness},
        ytick={0,1,2},
        legend style={at={(1.05,0.5)}, anchor=west, 
        minimum height=0.5cm,
        legend columns=1, column sep=0.2cm, draw=none, font=\footnotesize},
    ]
    \addplot+[mark=none,fill=5][error bars/.cd,x dir=both, x explicit]
        coordinates {
        (2.94,2) +- (0.09,0)
        (2.66,1) +- (0.32,0)
        (1.12,0) +- (0.09,0)
        };
    \addplot+[mark=none,fill=8][error bars/.cd,x dir=both, x explicit]
        coordinates { 
        (2.94,2) +- (0.11,0)
        (2.55,1) +- (0.46,0)
        (1.12,0) +- (0.16,0)
        };
    \addplot+[mark=none,fill=77][error bars/.cd,x dir=both, x explicit]
        coordinates { 
        (2.94,2) +- (0.11,0)
        (2.48,1) +- (0.46,0)
        (1.10,0) +- (0.10,0)
        };
    \addplot+[mark=none,fill=66][error bars/.cd,x dir=both, x explicit]
        coordinates {
        (2.83,2) +- (0.11,0)
        (2.49,1) +- (0.41,0)
        (1.18,0) +- (0.29,0)
        };    
    \legend{AICL, Few-shot, Zero-shot, Ground Truth}
    \end{axis}
    \end{tikzpicture}}
    \vspace{-0.15in}
    \caption{Human evaluation on inferred intents. The bar and horizontal line are mean and standard deviation values, respectively.}\label{fig:intent-human-evaluation}
    \vspace{-0.1in}
\end{figure*}

\subsection{Experimental Setup}
\subsubsection{Datasets}
We adopt three public bundle datasets created by a resource paper in SIGIR 2022~\cite{sun2022revisiting,sun2024revisiting}. In particular, 
they design a crowdsourcing task to annotate high-quality bundles and the corresponding intents from user sessions in Amazon datasets~\cite{he2016ups} with three domains, i.e., Electronic, Clothing, and Food. The statistics are summarized in Table~\ref{tab:datasets}. For each dataset, we chronologically split the session data into training, validation, and test sets with a ratio of 7:1:2. 
To the best of our knowledge, they are the ONLY bundle datasets with user sessions and well-labeled intents. Other widely-used bundle datasets, such as Steam, Netease, Youshu~\cite{avny2022bruce}, Goodreads~\cite{he2020consistency}, and iFashion~\cite{ren2023distillation} cannot be utilized in our study due to the unavailability of bundle intents.

\subsubsection{Baselines} We compare our proposed AICL with seven baselines. \textbf{Freq}~\cite{sun2022revisiting} is the frequent itemsets mining method. \textbf{BBPR}~\cite{pathak2017generating} is the greedy method with the predictions of BPRMF~\cite{rendle2012bpr}.
\textbf{POG}~\cite{chen2019pog} is the Transformer-based encoder-decoder model to generate personalized outfits.
\textbf{BYOB}~\cite{deng2021build} treats bundle generation as a combinatorial optimization problem with reinforcement learning.
\textbf{T5}~\cite{raffel2020exploring} is a Transformer-based seq2seq model.  
We use the version with 220M parameters and fine-tune it with our training data.
\textbf{Zero-shot} directly adopts LLMs to generate bundles and infer intents from user sessions. 
\textbf{Few-shot} exploits LLMs to perform the two tasks with few-shot examples. Furthermore, for a comprehensive comparison, we consider different variants for Few-shot by changing the way of selecting few-shot examples, including (1) \textit{Few-shot-random} randomly selecting different examples for each test session as the demonstration; (2) \textit{Few-shot-fix} randomly selecting the same examples, and use them as demonstrations for all test sessions; and (3) \textit{Few-shot-top} using the nearest neighbor sessions (same as in AICL) for each test session as the demonstration. We empirically find that Few-shot-fix generally achieves the best performance among all variants. Thus, we report the results produced by Few-shot-fix in our study. 
\textit{It is noteworthy that we do not compare with BUNT~\cite{he2022bundle}, Conna~\cite{wei2022towards} and BGGN~\cite{chang2021bundle}. This is because BUNT requires explicit user queries, and the source codes of Conna and BGGN are not available. We failed to reproduce them without the model details.}

\subsubsection{Evaluation Metrics} { Following~\cite{sun2022revisiting,sun2024revisiting}, we adopt three metrics to evaluate the quality of generated bundles: $Precision$, $Recall$, and $Coverage$.
At the session level, $Precision$ and $Recall$ measure how many bundles (subsets included) have been correctly predicted for each session. Meanwhile,  $Coverage$, at the bundle level, measures how many items are correctly covered by each hit bundle compared to the ground truth bundle. Due to space limitations, the detailed explanation of these metrics can be found in ~\cite{sun2022revisiting,sun2024revisiting}.}






Concerning the inferred intents, our initial plan was to perform an automatic evaluation using the widely-used ROUGE~\cite{rothe2020leveraging} to evaluate $n$-grams of the generated intents with ground truth. However, our empirical observations reveal that the generated intents, while semantically aligned with the ground truth, exhibit distinct expressions. Hence, using ROUGE may not accurately gauge and reflect the true quality of these intents.
Thus, we carefully design human evaluation to examine the quality of intents with the three metrics Naturalness, Coverage, and Motivation defined in Section~\ref{subsubsec:auto-feedback}. In our study, we ask 15 participants to rate the intents generated by the workers (ground truth) and three LLMs (Zero-shot, Few-shot, and AICL) for 60 bundles (20 per domain).

\subsubsection{Hyper-parameter Settings}
The best parameter settings for all methods are found based on the performance of the validation set or the suggested values in the original papers.
For Freq, we apply a grid search in $\{0.0001,0.001,0.01 \}$ for the \textit{support} and \textit{confidence} values. The best settings are 0.001 on all datasets. 
The embedding size is set to 20 for BBPR, POG, and BYOB for a fair comparison. The negative samples for BBPR and BYOB are set to 2. 
For BBPR, the initial bundle size is 3, and the number of neighbors is 10.
For BYOB, the bundle size is set as 3. 
For POG and BYOB, the size of candidate item set is 20.
The batch size for POG, BYOB, and T5 are set as 64, 64, and 4 respectively. The learning rate is searched in $\{0.0001,0.001,0.01\}$ for BBPR, POG, and BYOB, and in $\{0.00002, 0.00005, 0.00007, 0.0001\}$ to fine-tune T5. The best settings are 0.01, 0.001, 0.001, and 0.00005 for the four methods, respectively. 
For Zero- and Few-shot, we use the same prompts as AICL (bundle generation and intent reasoning parts only).
For Few-shot, we use one example to construct the demonstration. For a fair comparison, we set $k=1$ for our AICL. Besides, we apply a grid search in $\{1, 2, 3, 4, 5\}$ for $T_s, T_b$, and $T_i$. The optimal settings are $T_s=T_i=1$ and $T_b=4$. 

\subsection{Results and Analysis}

\subsubsection{Performance of Bundle Generation}
Table~\ref{tab:result-generation} shows the performance of all methods on the bundle generation task. 
Several interesting observations can be noted.
\textbf{(1)} LLM-based methods generally surpass Non-LLM ones, exhibiting the superiority of LLMs on our defined tasks. \textit{Regarding the Non-LLM methods}, \textbf{(2)} the straightforward Freq outperforms all model-based methods (BBPR, POG, and BYOB), possibly because the data sparsity issue causes the model-based methods to be under-trained. In contrast, Freq initially identifies frequent patterns at the category level, effectively mitigating such an issue. \textbf{(3)} Among the three model-based methods, the DL-based ones (POG and BYOB) exhibit better performance, underscoring the effectiveness of DL techniques.
\textit{In terms of LLM-based methods}, \textbf{(4)} T5 (220M) performs the least effectively, primarily due to its relatively small model size in comparison to GPT-3.5 with 154 billion parameters. \textbf{(5)} Few-shot exceeds Zero-shot, showcasing the usefulness of demonstrations in ICL. \textbf{(6)} AICL generally delivers the top performance across all datasets, providing solid evidence of the effectiveness and efficiency of its distinctive design.

\subsubsection{Performance of Intent Reasoning}
Figure~\ref{fig:intent-human-evaluation} displays the rating scores of intents generated by Zero-shot, Few-shot, AICL, and the workers (ground truth). Other baselines are not compared as they cannot generate intents. 
We randomly select 20 correctly generated bundles and their intents in each domain. In total, 60 bundles are assessed on three metrics, i.e., Naturalness, Coverage, and Motivation as defined in Section~\ref{subsubsec:auto-feedback}. 
The overall trends on all datasets are similar. First, AICL achieves the best performance in most cases, either with higher mean values or lower standard deviation values. This helps confirm the superiority of AICL on effective intent reasoning. Second, Few-shot generally exceeds Zero-shot, validating the usefulness of demonstrations on guiding LLMs for improved performance. Third, the ground truth intents annotated by workers are defeated by at least one of the three LLM-based methods (except `Motivation' on Food). 
This might be attributed to workers often prioritizing the speed of task completion to maximize their earnings, potentially at the expense of work quality~\cite{gadiraju2015understanding}.
Furthermore, it underscores the advanced capabilities of reasoning and natural language generation in LLMs, emphasizing their substantial potential in the context of crowdsourcing tasks.

\begin{table*}[!t]
\centering
\footnotesize
\addtolength{\tabcolsep}{6.8pt}
\caption{The results of ablation study on the bundle generation task. We run each variant five times to report the average results.}\label{tab:result-ablation}
\vspace{-0.15in}
\begin{tabular}{l|c|c|c|c|c|c|c|c|c}
\toprule
&\multicolumn{3}{c|}{Electronic} &\multicolumn{3}{c|}{Clothing} &\multicolumn{3}{c}{Food}  \\\cline{2-10}
&Precision&Recall&Coverage&Precision&Recall&Coverage&Precision&Recall&Coverage\\\midrule
AICL\_{w/o\_top} &0.667 &0.823 &0.734 &0.667 &0.761 &0.814 &0.689 &0.837 &0.743 \\ 
AICL\_{w/o\_self} &0.651 &0.839 &0.735 &0.649 &0.768 &0.835 &0.651 &0.820 &0.747 \\ 
AICL\_{w/o\_auto}  &0.635 &0.811 &0.719 &0.652 &0.768 &0.816 &0.663 &0.825 &0.729 \\ 
AICL\_{w/o\_context}  &0.556 &0.786 &0.721 &0.577 &0.723 &0.826 &0.583 &0.814 &0.723 \\ 
AICL\_{w/o\_rules} &0.665 &0.822 &0.729 &0.661 &0.772 &0.826 &0.679 &0.834 &0.742 \\ 
AICL\_{w/o\_intent} &0.636 &0.825 &0.731 &0.655 &0.769 &0.821 &0.671 &0.841 &0.741 \\  
AICL &\textbf{0.679} &\textbf{0.859} &\textbf{0.741} &\textbf{0.677} &\textbf{0.788} &\textbf{0.839} &\textbf{0.698} &\textbf{0.851} &\textbf{0.755} \\\bottomrule
\end{tabular}
\end{table*}

\begin{table}[t]
\centering
\footnotesize
\addtolength{\tabcolsep}{1pt}
\caption{The performance comparison between SentenceTransformer and LLMs on Electronic.}
\label{tab:sentencetransformer}
\vspace{-0.15in}
\begin{tabular}{l|ccc}
\toprule
& Precision &Recall &Coverage\\
\midrule
SentenceTransformer (all-MiniLM-L6-v2)  &0.623 &0.745 &0.674 \\
OpenAI (text-embedding-ada-002)	        &0.616 &0.860 &0.683 \\
\bottomrule   
\end{tabular}
\vspace{-0.1in}
\end{table}

\subsubsection{Ablation Study}\label{subsec:ablation}
We compare AICL with its six variants to examine the efficacy of each component. In particular, AICL\_{w/o\_top} randomly samples one session in the training set to replace the top neighbor. AICL\_{w/o\_self} removes the self-correction strategy from the demonstration. AICL\_{w/o\_auto} omits the auto-feedback mechanism from the demonstration. 
AICL\_{w/o\_context} removes both self-correction and auto-feedback modules from the demonstration.
AICL\_{w/o\_rules} abandons the rules summarization from the demonstration. 
AICL\_{w/o\_intent} deletes intent reasoning from the demonstration. Due to space limitations, we only show the results of the bundle generation task, as presented in Table~\ref{tab:result-ablation}. 

Overall, all the variants demonstrate lower performance compared to AICL, showcasing the contribution of each component to the improved performance.
To be specific, AICL\_{w/o\_top} underperforms AICL, which indicates the importance of identifying highly correlated examples to generate demonstrations. Both AICL\_{w/o\_self} and AICL\_{w/o\_auto} perform worse than AICL, while gaining better performance compared with AICL\_{w/o\_context}, implying the significance of both self-correction and auto-feedback strategies. The fact that AICL defeats AICL\_{w/o\_rules} exhibits the usefulness of rules summarization in instructing LLMs. Besides, an obvious performance drop is observed on AICL\_{w/o\_intent} when compared with AICL. This helps reinforce our claim that effective user intents play a crucial role in identifying relevant products to form improved bundles.

Furthermore, our Neighbor Session Retrieval module exploits the open-source Python framework SentenceTransformer (all-MiniLM-L6-v2) to get session representations for similarity calculation instead of using LLMs (e.g., OpenAI text-embedding-ada-002) due to its comparable performance with less cost (time and money) as explained in Section~\ref{subsec:nsr}. To verify our claim, we randomly sample 50 sessions from Electronic and use the two methods to help get the nearest neighbors. 
The results are presented in Table~\ref{tab:sentencetransformer}. Accordingly, their precision and coverage are comparable, while AICL with OpenAI embeddings possess higher recall. This also indicates the results of AICL reported in our paper may not be its upper bound, and there is still space for further improvements by using better sentence encoding models.

\pgfplotsset{
compat=1.11,
legend image code/.code={
\draw[mark repeat=2,mark phase=2]
plot coordinates {
(0cm,0cm)
(0.15cm,0cm)        
(0.3cm,0cm)         
};%
}
}
\begin{figure}[t]
	\centering
	\subfigure{
		\begin{tikzpicture}
		\begin{axis}[
		width=4.5cm,
		height=3.8cm,
		ylabel={Precision},
        xlabel={$T_b$},
		xmin=-0.5, xmax=5.5,
		ymin=0.49, ymax=0.66,
		xtick={0,1,2,3,4,5},
		yticklabel style={/pgf/number format/.cd,fixed,precision=3},
		ytick={0.50,0.55,0.60, 0.65},
		ylabel style = {font=\footnotesize},
        xlabel style = {font=\footnotesize},
		tick label style={font=\footnotesize},
		scaled ticks=false,
		legend style={at={(0.63,0.52)}, font=\footnotesize, anchor=north,legend columns=1, draw=none, fill=none,
        },
		]
	\addplot[color=8,
		solid,
		mark=*,
		mark options={solid},
		line width=1pt,
        mark size=1.2pt,
		smooth] coordinates { 
			(0, 0.526)
            (1, 0.543)
            (2, 0.588)
            (3, 0.599)
            (4, 0.614)
			(5, 0.578)
		};
        \addplot[ color=77, 
        solid,
		mark=square,
		mark options={solid},
		line width=1pt,mark size=1.5pt,
		smooth] coordinates { 
            (0, 0.528)
            (1, 0.543)
            (2, 0.579)
            (3, 0.581)
            (4, 0.595)
			(5, 0.559)
			};
	\addplot[ color=5, 
 	solid,
		mark=diamond,
		mark options={solid},
		line width=1pt,mark size=1.5pt,
		smooth] coordinates { 
            (0, 0.553)
            (1, 0.608)
            (2, 0.615)
            (3, 0.628)
            (4, 0.646)
			(5, 0.611)
        };
	\legend{Electronic, Clothing, Food}
	\end{axis}
	\end{tikzpicture}}
    \subfigure{
		\begin{tikzpicture}
		\begin{axis}[
		width=4.5cm,
		height=3.8cm,
		ylabel={Recall},
        y label style={at={(-0.26,0.5)}},
        xlabel={$T_b$},
		xmin=-0.5, xmax=5.5,
		ymin=0.64, ymax=0.91,
		xtick={0,1,2,3,4,5},
		yticklabel style={/pgf/number format/.cd,fixed,precision=3},
        ytick={0.65, 0.75, 0.85},
		ylabel style = {font=\footnotesize},
        xlabel style = {font=\footnotesize},
		tick label style={font=\footnotesize},
		scaled ticks=false,
		legend style={at={(0.6,0.82)}, font=\footnotesize, anchor=north,legend columns=1, draw=none, fill=none,},
		]
	\addplot[color=8,
		solid,
		mark=*,
		mark options={solid},
		line width=1pt,
        mark size=1.2pt,
		smooth] coordinates { 
            (0, 0.775)
			(1, 0.832)
            (2, 0.870)
            (3, 0.900)
            (4, 0.897)
			(5, 0.832)
		};
    \addplot[color=77,
		solid,
		mark=square,
		mark options={solid},
		line width=1pt,mark size=1.5pt,
		smooth] coordinates { 
            (0, 0.669)
			(1, 0.671)
            (2, 0.679)
            (3, 0.687)
            (4, 0.691)
			(5, 0.673)
		};
    \addplot[color=5,
		solid,
		mark=diamond,
		mark options={solid},
		line width=1pt,mark size=1.5pt,
		smooth] coordinates { 
            (0, 0.698)
			(1, 0.711)
            (2, 0.713)
            (3, 0.721)
            (4, 0.731)
			(5, 0.709)
		};
    \legend{Electronic, Clothing, Food}
	\end{axis}
	\end{tikzpicture}}
	\vspace{-0.2in}
	\caption{The impact of $T_b$ on all neighbor sessions.}
\label{fig:hyper-parameter}
	\vspace{-0.15in}
\end{figure}
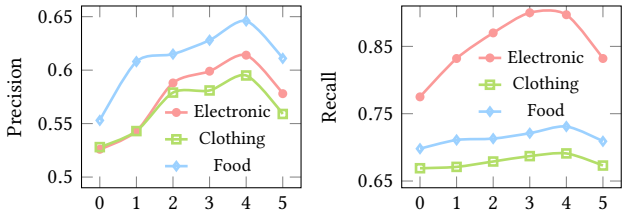

\subsubsection{Hyper-Parameter Analysis} We further study the impact of essential hyper-parameters on AICL, including the rounds of self-correction ($T_s$) and auto-feedback for both tasks ($T_b$ and $T_i$), as well as the number of neighbor sessions ($k$). 
First, we observe that around 29\% of neighbor sessions adjust bundles and intents with $T_s=1$, and the accuracy is improved by 5.9\% w.r.t. Precision on average. In summary, the self-correction allows LLMs to reassess the response, leading to more self-consistent and effective results.
Second, we apply a grid search in $\{1, 2, 3, 4, 5\}$ to check the impact of $T_b$ depicted in Figure~\ref{fig:hyper-parameter}. 
As $T_b$ increases, the performance initially rises, reaching its peak with $T_b=4$, and subsequently declines as $T_b$ continues to increase. 
Our empirical findings indicate an average improvement of 16.7\% in Precision with the auto-feedback.
Third, we find that with $T_i=1$, most intents (65.7\%) generated by LLMs are quite close to the ground truth intents annotated by workers. It reveals the great potential of LLMs in crowdsourcing tasks. 
Lastly, we observe that the best performance is attained with $k=1$. 
Increasing the value of $k$ does not consistently yield noticeable improvements and can, on occasion, even lead to marginal performance declines. This is intuitive as a large $k$ substantially lengthens the context, which may confuse LLMs and result in decreased performance.

\subsubsection{Case Study}
\begin{figure}[t]
\centering
\includegraphics[width=0.98\linewidth]{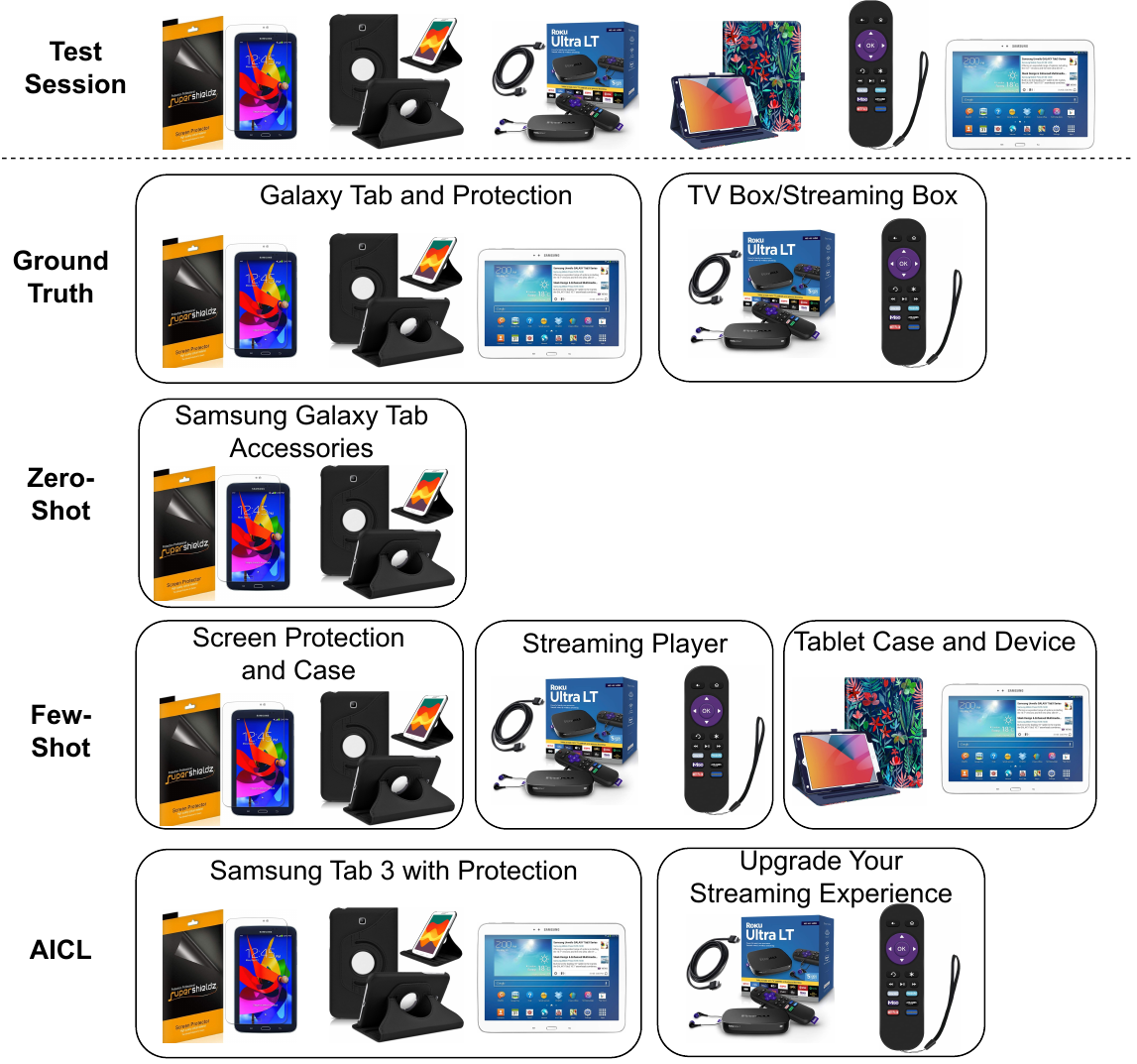}
\vspace{-0.05in}
\caption{The generated bundles and intents on Electronic.}
\label{fig:good-case}
\vspace{-0.05in}
\end{figure}

A case study is performed to check the generated bundles and intents by Zero-shot, Few-shot, and AICL. Due to space limitations, we only show one sampled test session on Electronic in Figure~\ref{fig:good-case}.    
For \textit{bundle generation}, it's evident that Zero-shot only manages to generate a portion of the Galaxy Tab and Protection bundle, overlooking the TV Box bundle. Few-shot, on the other hand, identifies three bundles, but the first is a subset of the ground truth, and the last, combining the iPad case and Galaxy Tab, is less coherent. In contrast, AICL consistently and effectively generates all bundles that perfectly align with the ground truth.   
For \textit{intent reasoning}, all methods perform similarly in terms of Naturalness and Coverage. However, AICL demonstrates a superior performance in terms of Motivation. For example, the intent associated with the TV Box bundle `Upgrade Your Streaming Experience' is more motivational than the intents `Streaming Box' (ground truth) and `Streaming Player' (Few-shot).

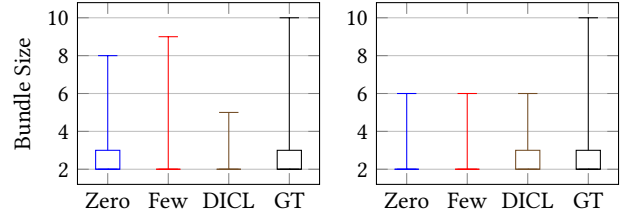
\begin{figure}[t]
    \subfigure{
    \begin{tikzpicture}
\begin{axis}[
   boxplot/draw direction=y,
   ymajorgrids,
   ylabel={Bundle Size},
   xtick={1, 2, 3, 4},
   xticklabels={Zero, Few, AICL, GT},
   xmin=0.5,
   xmax=4.5,
   width=4.8cm,
   height=3.8cm,
   ylabel style ={font = \footnotesize},
   xlabel style ={font = \footnotesize},
   tick label style={font=\footnotesize}
]
\addplot+[boxplot prepared={
      draw position=1,
      lower whisker=2,
      lower quartile=2,
      median=2,
      upper quartile=3,
      upper whisker=8,
      box extend=0.4,
      },] coordinates{};  
\addplot+[boxplot prepared={
      draw position=2,
      lower whisker=2,
      lower quartile=2,
      median=2,
      upper quartile=2,
      upper whisker=9,
      box extend=0.4,
      },] coordinates{};
\addplot+[boxplot prepared={
      draw position=3,
      lower whisker=2,
      lower quartile=2,
      median=2,
      upper quartile=2,
      upper whisker=5,
      box extend=0.4,
      },] coordinates{}; 
\addplot+[boxplot prepared={
      draw position=4,
      lower whisker=2,
      lower quartile=2,
      median=2,
      upper quartile=3,
      upper whisker=10,
      box extend=0.4,
      },] coordinates{}; 
\end{axis}
\end{tikzpicture}
}
\subfigure{
\begin{tikzpicture}
\begin{axis}[
   boxplot/draw direction=y,
   ymajorgrids,
   xmin=0.5,
   xmax=4.5,
   xtick={1, 2, 3, 4},
   xticklabels={Zero, Few, AICL, GT},
   width=4.8cm,
   height=3.8cm,
   tick label style={font=\footnotesize}
]
\addplot+[boxplot prepared={
      draw position=1,
      lower whisker=2,
      lower quartile=2,
      median=2,
      upper quartile=2,
      upper whisker=6,
      box extend=0.4,
      },] coordinates{};   
\addplot+[boxplot prepared={
      draw position=2,
      lower whisker=2,
      lower quartile=2,
      median=2,
      upper quartile=2,
      upper whisker=6,
      box extend=0.4,
      },] coordinates{};  
\addplot+[boxplot prepared={
      draw position=3,
      lower whisker=2,
      lower quartile=2,
      median=2,
      upper quartile=3,
      upper whisker=6,
      box extend=0.4,
      },] coordinates{};
\addplot+[boxplot prepared={
      draw position=4,
      lower whisker=2,
      lower quartile=2,
      median=2,
      upper quartile=3,
      upper whisker=10,
      box extend=0.4,
      },] coordinates{}; 
\end{axis}
\end{tikzpicture}
    }
\vspace{-0.15in}
\caption{Bundle size distribution on Electronic and Clothing.}\label{fig:bundle-size}
\end{figure}

\subsubsection{Limitations of LLMs for Bundle Generation}
\begin{figure}[t]
\centering
\includegraphics[width=0.98\linewidth]{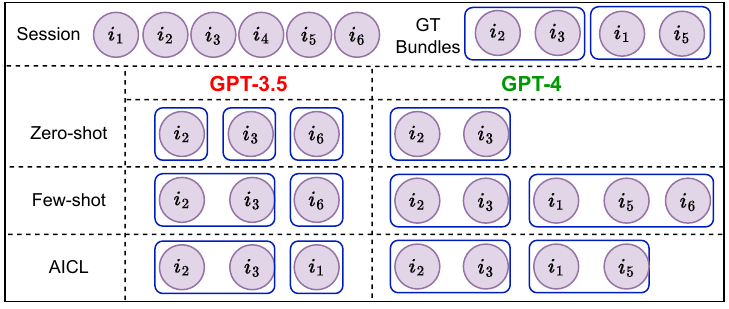}
\vspace{-0.1in}
\caption{The comparison between GPT-3.5 and GPT-4.}
\label{fig:bad-case}
\vspace{-0.05in}
\end{figure}

Despite the effectiveness of LLMs, we now discuss their limitations on our defined tasks. First, through empirical observations, we notice that LLM-based methods tend to produce smaller bundles in comparison to the ground truth (GT) bundles annotated by workers, as depicted in Figure~\ref{fig:bundle-size}. 
The smaller bundle size could limit the diversity of recommendations. However, it may align with real user behavior, as most online customers often purchase a small number, typically two, of items in one shopping session~\cite{zhu2014bundle}. In this context, an overlarge bundle size may not provide significant benefits and could potentially divert users' attention.
Second, despite our various attempts to emphasize the prompt constraint, namely, `Each bundle must contain multiple products', GPT-3.5 sometimes generates bundles with only a single product. It could be attributable to its inherent hallucination issue, and such an issue can be partially resolved by more powerful LLMs, e.g., GPT-4. Specifically, we do a preliminary exploration to examine the performance of GPT-4 on sessions (5 per domain) with such an issue using GPT-3.5. The results are depicted in Table~\ref{tab:result-gpt4}. Overall, improvements are observed in two key areas. Firstly, all compared methods exhibited fewer instances of `bad cases' (i.e., bundles containing a single product) with GPT-4 compared to GPT-3.5. Secondly, the accuracy (precision, recall, and coverage) of most methods show enhancements with GPT-4 compared to GPT-3.5, although there are some exceptions, such as decreases in certain metrics. Moreover, Figure~\ref{fig:bad-case} illustrates the performance comparison on a real test session on Electronic, which has such a hallucination issue with GPT-3.5 but can be completely resolved with GPT-4.  

\begin{table}[t]
\centering
\footnotesize
\addtolength{\tabcolsep}{-1.5pt}
\caption{The performance comparison between GPT-3.5 and GPT-4, where `\#Bad Case' refers to the number of sessions with the hallucination issue. Due to space limitation, we only present the results on Precision and Recall.  
}\label{tab:result-gpt4}
\vspace{-0.15in}
\begin{tabular}{l|l|ccc|ccc}
\toprule
\multicolumn{2}{l|}{} &\multicolumn{3}{c|}{\textbf{GPT-3.5}}
&\multicolumn{3}{c}{\textbf{GPT-4}} \\\cline{3-8}
\multicolumn{2}{l|}{} &Recall &Precision &\#Bad Case &Recall &Precision &\#Bad Case\\
\midrule
\multirow{3}{*}{\rotatebox[origin=c]{90}{Elect.}} &Zero-shot
&0.733 &0.667 &5 &0.733 &0.633 &2\\
&Few-shot
&0.533 &0.600 &5 &0.800 &0.600 &1\\
&AICL
&0.800 &0.700 &5 &0.833 &0.700 &1\\\hline
\multirow{3}{*}{\rotatebox[origin=c]{90}{Clothing}} &Zero-shot
&0.533 &0.517 &5 &0.567 &0.450 &2\\
&Few-shot
&0.433 &0.400 &5 &0.667 &0.483 &1\\
&AICL
&0.611 &0.556 &5 &0.667 &0.583 &1\\\hline
\multirow{3}{*}{\rotatebox[origin=c]{90}{Food}} &Zero-shot
&0.800 &0.733 &5 &0.800 &0.667 &2\\
&Few-shot
&0.646 &0.533 &5 &0.800 &0.667 &2\\
&AICL
&0.800 &0.733 &5 &0.800 &0.700 &1\\
\bottomrule
\end{tabular}
\end{table}

\section{Conclusion and Future Work}
Motivated by the advanced reasoning capability exhibited in LLMs, this paper initiates a pioneering exploration into two interrelated tasks, i.e., personalized bundle generation and the underlying intent inference, both rooted in users' behaviors within a session.
To this end, we propose an adaptive in-context learning (AICL) paradigm equipped with three modules, i.e., neighbor session retrieval, adaptive demonstration generation, and demonstration guided inference. This empowers LLMs to seek tailored and adaptive lessons from neighbor sessions as demonstrations for performance improvements on the test session. As a result, it ultimately delivers an effective approach that is capable of generating multiple, personalized, and intelligible bundles with adaptive sizes given any user session. Experimental results on three public datasets verify the effectiveness of our AICL on both tasks.

For future endeavors, there are several potential directions, including (1) designing strategies to create larger bundles and better control the output format; 
(2) introducing self-correction and auto-feedback mechanisms in the inference stage, and (3) exploring the utilization of multi-modal data for further enhancement.

\begin{acks}
We greatly acknowledge the support of National Natural Science Foundation of China (Grant No. 72371148 and 72192832), the Shanghai Rising-Star Program (Grant No. 23QA1403100), and the Natural Science Foundation of Shanghai (Grant No. 21ZR1421900). It was also supported by A*Star Center for Frontier Artificial Intelligence Research and in part by the Data Science and Artificial Intelligence Research Centre, School of Computer Science and Engineering at the Nanyang Technological University (NTU), Singapore.
\end{acks}

\clearpage
\newpage
\bibliographystyle{ACM-Reference-Format}
\balance
\bibliography{reference}

\end{document}